\documentclass[aps,prd,superscriptaddress,nofootinbib,preprintnumbers,notitlepage]{revtex4-1}

\usepackage{graphicx}
\usepackage{dcolumn}
\usepackage{bm}
\usepackage{amssymb,amsmath,epsfig}
\usepackage{color}
\usepackage{slashed}
\usepackage{verbatim}
\usepackage{multirow}
\usepackage{braket}

\allowdisplaybreaks

\newcommand{\p}{\partial}
\newcommand{\st}{{\scriptscriptstyle T}}
\def\nn{\nonumber}
\def\cd{{\cdot}}

\usepackage[normalem]{ulem} 
\renewcommand\sout{\bgroup \color[rgb]{0.55,0.00,0.99} \ULdepth=-.5ex \ULset}

\begin{document}

\title{Quark and gluon TMD correlators in momentum and coordinate space}
\author{Tom van Daal}
\email{tvdaal@nikhef.nl}
\affiliation{Department of Physics and Astronomy, VU University Amsterdam, De Boelelaan 1081, NL-1081 HV Amsterdam, The Netherlands}
\affiliation{Nikhef, Science Park 105, NL-1098 XG Amsterdam, The Netherlands}

\date{\today}

\preprint{NIKHEF 2016-060}

\begin{abstract}
Transverse momentum dependent (TMD) distribution correlators can be parametrized in terms of TMD parton distribution functions (PDFs), or TMDs for short. We provide an overview of the leading-twist quark and gluon TMDs, both in momentum space and in coordinate space (also called $b_\st$-space, where $b_\st$ is Fourier conjugate to the partonic momentum $k_\st$). We consider unpolarized, vector polarized, as well as tensor polarized hadrons, which is relevant for spin-$0$, spin-$1/2$, and spin-$1$ hadrons.
\end{abstract}

\maketitle

\section{Introduction} \label{s:intro}
The extraction of a parton from a hadron is mathematically described by a distribution correlator. A correlator cannot be calculated using perturbative quantum chromodynamics (QCD). Rather, it is parametrized in terms of parton distribution functions (PDFs). The so-called collinear PDFs are probability densities in the longitudinal momentum fraction $x$ of the parton with respect to the parent hadron. Going beyond a collinear treatment, one can define transverse momentum dependent (TMD) PDFs, or TMDs for short, that besides $x$ also depend on the partonic transverse momentum $k_\st$. The TMDs encode the three-dimensional inner structure of hadrons. Employing the lightlike vectors $\bar{n}^\mu \equiv [1,0,\bm{0}_\st]$ and $n^\mu \equiv [0,1,\bm{0}_\st]$, the hadron and parton momenta $P$ and $k$ are given by:\footnote{We define the light-cone components of a four-vector $a$ in terms of its Minkowski components as $a^\pm \equiv (a^0 \pm a^3)/\sqrt{2}$ (the transverse components are unchanged). Using light-cone coordinates, we can represent a four-vector as $a^\mu = [a^+,a^-,\bm{a}_\st]$. We also define $a_\st^\mu \equiv [0,0,\bm{a}_\st]$, so that $a_\st^2 = - \bm{a}_\st^2$.}
\begin{align}
    P^\mu &= P^+ \bar{n}^\mu + \frac{M^2}{2P^+} \,n^\mu , \\[5pt]
    k^\mu &= xP^+ \bar{n}^\mu + k^- n^\mu + k_\st^\mu ,
\end{align}
where $M$ is the mass of the hadron.

The inclusion of hadron polarization forms a key component of TMD phenomenology. An ensemble of spin-$1/2$ hadrons can be unpolarized or vector polarized. To describe the degree of vector polarization, we make use of a spin vector $S$. For spin-$1$ hadrons we need besides $S$ also a symmetric traceless spin tensor $T$ to describe tensor polarized states. Satisfying $P \cd S = 0$ and $P_\mu T^{\mu\nu} = 0$, the spin vector and tensor can be parametrized as follows~\cite{Bacchetta:2000jk}:
\begin{align}
    S^\mu &= S_L \frac{P^+}{M} \,\bar{n}^\mu - S_L \frac{M}{2 P^+} \,n^\mu + S_T^\mu , \\[5pt]
    T^{\mu\nu} &= \frac{1}{2} \left[ \frac{4}{3} S_{LL} \frac{(P^+)^2}{M^2} \,\bar{n}^\mu \bar{n}^\nu + \frac{P^+}{M} \,\bar{n}_{\phantom{LT}}^{\{\mu} \!\!S_{LT}^{\nu\}} \right. \nn \\
    &\left. \quad\quad\;\; - \,\frac{2}{3} S_{LL} \left( \bar{n}^{\{\mu} n^{\nu\}} - g_\st^{\mu\nu} \right) + S_{TT}^{\mu\nu} \right. \nn \\
    &\left. \quad\quad\;\; - \,\frac{M}{2P^+} \,n_{\phantom{LT}}^{\{\mu} \!\!S_{LT}^{\nu\}} + \frac{1}{3} S_{LL} \frac{M^2}{(P^+)^2} \,n^\mu n^\nu \right] ,
\end{align}
where curly brackets denote symmetrization of indices, and $g_\st^{\mu\nu} \equiv g^{\mu\nu} - \bar{n}^{\{\mu} n^{\nu\}}$ with nonzero components $g_\st^{11} = g_\st^{22} = -1$. The spin vector has three independent parameters, namely $S_L$ and the two transverse components of $S_T$, whereas the spin tensor has five, namely $S_{LL}$, the two transverse components of $S_{LT}$, as well as the two independent components of the symmetric traceless transverse tensor $S_{TT}$.

Parametrizations of hadronic correlators in terms of TMDs are usually given in momentum space (or $k_\st$-space). However, for certain applications such as the implementation of TMD evolution, correlators are studied in coordinate space (or $b_\st$-space, where $b_\st$ is Fourier conjugate to $k_\st$). To ensure a one-to-one correspondence between TMDs in $k_\st$-space and $b_\st$-space, it is essential that a correlator is parametrized in terms of TMDs of definite rank. To this end, a correlator in $k_\st$-space needs to be parametrized using symmetric traceless tensors in $k_\st$~\cite{Boer:2016xqr}. Up to rank $n=4$, the symmetric traceless tensors $k_\st^{i_1 \ldots i_n}$ are given by
\begin{align}
    k_\st^{ij} \equiv& \;k_\st^i k_\st^j + \frac{1}{2} \bm{k}_\st^2 g_\st^{ij} , \\
    k_\st^{ijk} \equiv& \;k_\st^i k_\st^j k_\st^k + \frac{1}{4} \bm{k}_\st^2 \left(
    g_\st^{ij} k_\st^k + g_\st^{ik} k_\st^j + g_\st^{jk} k_\st^i \right) , 
    \label{kT3}\\
    k_\st^{ijkl} \equiv& \;k_\st^i k_\st^j k_\st^k k_\st^l + \frac{1}{6} \bm{k}_\st^2 \left( g_\st^{ij} k_\st^{kl} + g_\st^{ik} k_\st^{jl} + g_\st^{il} k_\st^{jk} + g_\st^{jk} k_\st^{il} + g_\st^{jl} k_\st^{ik} + g_\st^{kl} k_\st^{ij} \right) \nn \\
    & - \frac{1}{8} \bm{k}_\st^4 \left( g_\st^{ij} g_\st^{kl} + g_\st^{ik} g_\st^{jl} + g_\st^{il} g_\st^{jk} \right) ,
\end{align}
satisfying
\begin{equation}
    {g_{\st}}_{ij} k_\st^{ij} = {g_{\st}}_{ij} k_\st^{ijk} = {g_{\st}}_{ij} k_\st^{ijkl} = 0 .
\end{equation}
See appendix C of ref.~\cite{Boer:2016xqr} for useful identities involving products of these tensors. The symmetric traceless tensor $k_\st^{i_1 \ldots i_n}$ of rank $n \geq 1$ only has two independent components, which allows for the following decomposition:
\begin{equation}
    k_\st^{i_1 \ldots i_n} \;\to\; \frac{|\bm{k}_\st|^n}{2^{n-1}} \,e^{\pm in\varphi} ,
    \label{e:STT} 
\end{equation}
in terms of the two real numbers $|\bm{k}_\st|$ and $\varphi$, the polar coordinates of the transverse vector $k_\st$.

\section{Correlators in momentum space}
\subsection{The quark-quark TMD correlator}
The quark TMDs that appear in the parametrization of the quark-quark correlator were introduced in the nineties in refs.~\cite{Tangerman:1994eh,Boer:1997nt} for unpolarized and vector polarized hadrons, and in 2000 also tensor polarized hadrons were considered~\cite{Bacchetta:2000jk}. For the quark TMDs the notation of ref.~\cite{Bacchetta:2000jk} is used here, which also coincides with the notation in refs.~\cite{Tangerman:1994eh,Boer:1997nt} for the unpolarized and vector polarized cases. The quark-quark TMD correlator for spin-$1$ hadrons is given by\footnote{To obtain the correlator for a spin-$0$ or spin-$1/2$ hadron, one simply sets $S = T = 0$ or $T = 0$ respectively.}
\begin{align}
    \Phi_{ij}(x,\bm{k}_\st) \equiv& \int \left. \frac{d\xi^- d^2\xi_\st}{(2\pi)^3} \,e^{ik\cdot\xi} \bra{P,S,T} \overline{\psi}_j(0) \,U_{[0,\xi]} \,\psi_i(\xi) \ket{P,S,T} \vphantom{\int} \right|_{\xi^+=0} ,
    \label{e:quarkcorrelator}
\end{align}
where a summation over color is implicitly assumed. The gauge link $U_{[0,\xi]}$ is needed for color gauge invariance and gives rise to a process dependence of the TMDs.

The correlator in eq.~\eqref{e:quarkcorrelator} can be parametrized in terms of TMDs. To establish a one-to-one correspondence between the TMDs in momentum and coordinate space, we employ symmetric traceless tensors built from $k_\st$ (see section~\ref{s:intro}). This was already done in ref.~\cite{Boer:2016xqr} for the case of gluons, and in this proceedings contribution the same is done for the quark case as well. Separating the various possible hadronic polarization states, the correlator in eq.~\eqref{e:quarkcorrelator} can be parametrized in terms of leading-twist quark TMDs of definite rank as follows:
\begin{align}
    \Phi(x,\bm{k}_\st) =& \;\Phi_U(x,\bm{k}_\st) + \Phi_L(x,\bm{k}_\st) + \Phi_T(x,\bm{k}_\st) + \Phi_{LL}(x,\bm{k}_\st) + \Phi_{LT}(x,\bm{k}_\st) + \Phi_{TT}(x,\bm{k}_\st) ,
    \label{e:phiquark}
\end{align}
where\footnote{To avoid clutter, we suppress in this proceedings contribution in the names of the functions a reference to quarks, gluons, or gauge links.}
\begin{align}
    \Phi_U(x,\bm{k}_\st) &= \frac{1}{2} \left[ \slashed{\bar{n}} \,f_1(x,\bm{k}_\st^2) + \frac{\sigma_{\mu\nu} k_\st^\mu \bar{n}^\nu}{M} \,h_1^\perp(x,\bm{k}_\st^2) \right] , \\[5pt]
    \Phi_L(x,\bm{k}_\st) &= \frac{1}{2} \left[ \gamma^5 \slashed{\bar{n}} \,S_L \,g_1(x,\bm{k}_\st^2) + \frac{i\sigma_{\mu\nu} \gamma^5 \bar{n}^\mu k_\st^\nu S_L}{M} \,h_{1L}^\perp(x,\bm{k}_\st^2) \right] , \\[5pt]
    \Phi_T(x,\bm{k}_\st) &= \frac{1}{2} \left[ \frac{\slashed{\bar{n}} \,\epsilon_\st^{S_\st k_\st}}{M} \,f_{1T}^\perp(x,\bm{k}_\st^2) + \frac{\gamma^5 \slashed{\bar{n}} \,\bm{k}_\st \cd \bm{S}_\st}{M} \,g_{1T}(x,\bm{k}_\st^2) \right. \nn \\
    &\quad\quad\;\, \left. + \,i\sigma_{\mu\nu} \gamma^5 \bar{n}^\mu S_\st^\nu \,h_1(x,\bm{k}_\st^2) - \frac{i\sigma_{\mu\nu} \gamma^5 \bar{n}^\mu k_\st^{\nu\rho} {S_\st}_\rho}{M^2} \,h_{1T}^\perp(x,\bm{k}_\st^2) \right] , \\[5pt]
    \Phi_{LL}(x,\bm{k}_\st) &= \frac{1}{2} \left[ \slashed{\bar{n}} \,S_{LL} \,f_{1LL}(x,\bm{k}_\st^2) + \frac{\sigma_{\mu\nu} k_\st^\mu \bar{n}^\nu S_{LL}}{M} \,h_{1LL}^\perp(x,\bm{k}_\st^2) \right] , \\[5pt]
    \Phi_{LT}(x,\bm{k}_\st) &= \frac{1}{2} \left[ \frac{\slashed{\bar{n}} \,\bm{k}_\st \cd \bm{S}_{LT}}{M} \,f_{1LT}(x,\bm{k}_\st^2) + \frac{\gamma^5 \slashed{\bar{n}} \,\epsilon_\st^{S_{LT}k_\st}}{M} \,g_{1LT}(x,\bm{k}_\st^2) \right. \nn \\
    &\quad\quad\;\; \left. + \,\sigma_{\mu\nu} \bar{n}^\nu S_{LT}^\mu \,h_{1LT}(x,\bm{k}_\st^2) - \frac{\sigma_{\mu\nu} \bar{n}^\nu k_\st^{\mu\rho} {S_{LT}}_\rho}{M^2} \,h_{1LT}^\perp(x,\bm{k}_\st^2) \right] , \\[5pt]
    \Phi_{TT}(x,\bm{k}_\st) &= \frac{1}{2} \left[ \frac{\slashed{\bar{n}} \,k_\st^{\mu\nu} {S_{TT}}_{\mu\nu}}{M^2} \,f_{1TT}(x,\bm{k}_\st^2) - \frac{\gamma^5 \slashed{\bar{n}} \,{\epsilon_\st}_{\mu\nu} k_\st^{\mu\rho} {S_{TT}^\nu}_\rho}{M^2} \,g_{1TT}(x,\bm{k}_\st^2) \right. \nn \\
    &\quad\quad\;\; \left. - \,\frac{\sigma_{\mu\nu} \bar{n}^\nu k_\st^\rho {S_{TT}^\mu}_\rho}{M} \,h_{1TT}(x,\bm{k}_\st^2) + \frac{\sigma_{\mu\nu} \bar{n}^\nu k_\st^{\mu\rho\sigma} {S_{TT}}_{\rho\sigma}}{M^3} \,h_{1TT}^\perp(x,\bm{k}_\st^2) \right] ,
\end{align}
where we have employed the notation $\epsilon_\st^{ab} \equiv \epsilon_\st^{\mu\nu} a_\mu b_\nu$, with $\epsilon_\st^{\mu\nu} \equiv \epsilon^{\mu\nu-+}$ (the nonzero components are $\epsilon_\st^{12} = - \epsilon_\st^{21} = 1$). 

The functions $h_{1T}$, $h_{1LT}'$, and $h_{1TT}'$ that appear in the original parametrizations in refs.~\cite{Tangerman:1994eh,Bacchetta:2000jk} have been replaced by\footnote{These definitions were already proposed in ref.~\cite{Bacchetta:2000jk}, so there is no conflict of notation.}
\begin{align}
    h_1(x,\bm{k}_\st^2) &\equiv h_{1T}(x,\bm{k}_\st^2) + \frac{\bm{k}_\st^2}{2M^2} \,h_{1T}^\perp(x,\bm{k}_\st^2) , \\
    h_{1LT}(x,\bm{k}_\st^2) &\equiv h_{1LT}'(x,\bm{k}_\st^2) + \frac{\bm{k}_\st^2}{2M^2} \,h_{1LT}^\perp(x,\bm{k}_\st^2) , \\
    h_{1TT}(x,\bm{k}_\st^2) &\equiv h_{1TT}'(x,\bm{k}_\st^2) + \frac{\bm{k}_\st^2}{2M^2} \,h_{1TT}^\perp(x,\bm{k}_\st^2) .
\end{align}
Furthermore, we have defined $g_1 \equiv g_{1L}$.

\subsection{The gluon-gluon TMD correlator}
The gluon-gluon TMD correlator can be parametrized in terms of gluon TMDs. They were first introduced for unpolarized and vector polarized hadrons in 2001 in ref.~\cite{Mulders:2000sh}. A new nomenclature was proposed a couple of years later in ref.~\cite{Meissner:2007rx}. This year, also the case of tensor polarized hadrons has been considered~\cite{Boer:2016xqr}. The gluon-gluon TMD correlator for spin-$1$ hadrons is given by
\begin{align}
    \Gamma^{\mu\nu;\rho\sigma}(x,\bm{k}_\st) \equiv& \int \left. \frac{d\xi^- d^2\xi_\st}{(2\pi)^3} \,e^{ik\cdot\xi} \bra{P,S,T} F^{\mu\nu}(0) \,U_{[0,\xi]} \,F^{\rho\sigma}(\xi) \,U'_{[\xi,0]} \ket{P,S,T} \right|_{\xi^+=0} ,
    \label{e:gluoncorrelator}
\end{align}
where a trace in color space is implicitly assumed. Two process-dependent gauge links $U_{[0,\xi]}$ and $U'_{[\xi,0]}$ are needed for color gauge invariance. 

At leading twist, the gluon-gluon TMD correlator is given by $\Gamma^{ij}(x,\bm{k}_\st) \equiv \Gamma^{+i;+j}(x,\bm{k}_\st)$, where $i,j$ are transverse indices. Separating the various possible hadronic polarization states, this correlator can be parametrized in terms of leading-twist gluon TMDs of definite rank as follows~\cite{Boer:2016xqr}:
\begin{align}
    \Gamma^{ij}(x,\bm{k}_\st) =& \;\Gamma_U^{ij}(x,\bm{k}_\st) + \Gamma_L^{ij}(x,\bm{k}_\st) + \Gamma_T^{ij}(x,\bm{k}_\st) + \Gamma_{LL}^{ij}(x,\bm{k}_\st) + \Gamma_{LT}^{ij}(x,\bm{k}_\st) + \Gamma_{TT}^{ij}(x,\bm{k}_\st) ,
    \label{e:gammagluon}
\end{align}
where
\begin{align}
    \Gamma_U^{ij}(x,\bm{k}_\st) &= \frac{xP^+}{2} \left[ - \,g_\st^{ij} \,f_1(x,\bm{k}_\st^2) + \frac{k_\st^{ij}}{M^2} \,h_1^{\perp}(x,\bm{k}_\st^2) \right] , \\[5pt]
    \Gamma_L^{ij}(x,\bm{k}_\st) &= \frac{xP^+}{2} \left[ i \epsilon_\st^{ij} S_L \,g_1(x,\bm{k}_\st^2) + \frac{{\epsilon_\st^{\{i}}_\alpha k_\st^{j\}\alpha} S_L}{2M^2} \,h_{1L}^\perp(x,\bm{k}_\st^2) \right] , \\[5pt]
    \Gamma_T^{ij}(x,\bm{k}_\st) &= \frac{xP^+}{2} \left[ - \,\frac{g_\st^{ij} \epsilon_\st^{S_\st k_\st}}{M} \,f_{1T}^\perp(x,\bm{k}_\st^2) + \frac{i \epsilon_\st^{ij} \bm{k}_\st \cd \bm{S}_\st}{M} \,g_{1T}(x,\bm{k}_\st^2) \right. \nn \\
    &\qquad\qquad\, \left. - \,\frac{\epsilon_\st^{k_\st\{i} S_\st^{j\}} + \epsilon_\st^{S_\st\{i} k_\st^{j\}}}{4M} \,h_1(x,\bm{k}_\st^2) - \frac{{\epsilon_\st^{\{i}}_\alpha k_\st^{j\}\alpha S_\st}}{2M^3} \,h_{1T}^\perp(x,\bm{k}_\st^2) \right] , \\[5pt]
    \Gamma_{LL}^{ij}(x,\bm{k}_\st) &= \frac{xP^+}{2} \left[ - \,g_\st^{ij} S_{LL} \,f_{1LL}(x,\bm{k}_\st^2) + \frac{k_\st^{ij} S_{LL}}{M^2} \,h_{1LL}^{\perp}(x,\bm{k}_\st^2) \right] , \\[5pt]
    \Gamma_{LT}^{ij}(x,\bm{k}_\st) &= \frac{xP^+}{2} \left[ - \,\frac{g_\st^{ij} \bm{k}_\st \cd \bm{S}_{LT}}{M} \,f_{1LT}(x,\bm{k}_\st^2) + \frac{i \epsilon_\st^{ij} \epsilon_\st^{S_{LT}k_\st}}{M} \,g_{1LT}(x,\bm{k}_\st^2) \right. \nn \\
    &\qquad\qquad\, \left. + \,\frac{S_{LT}^{\{i} k_\st^{j\}}}{M} \,h_{1LT}(x,\bm{k}_\st^2) + \frac{k_\st^{ij\alpha} {S_{LT}}_\alpha}{M^3} \,h_{1LT}^{\perp}(x,\bm{k}_\st^2) \right] , \\[5pt]
    \Gamma_{TT}^{ij}(x,\bm{k}_\st) &= \frac{xP^+}{2} \left[ - \,\frac{g_\st^{ij} k_\st^{\alpha\beta} S_{TT\alpha\beta}}{M^2} \,f_{1TT}(x,\bm{k}_\st^2) + \frac{i \epsilon_\st^{ij} {\epsilon^{\beta}_\st}_\gamma k_\st^{\gamma\alpha} S_{TT\alpha\beta}}{M^2} \,g_{1TT}(x,\bm{k}_\st^2) \right. \nn \\
    &\qquad\qquad\, \left. + \,S_{TT}^{ij} \,h_{1TT}(x,\bm{k}_\st^2) + \frac{{S_{TT}^{\{i}}_\alpha k_\st^{j\}\alpha}}{M^2} \,h_{1TT}^{\perp}(x,\bm{k}_\st^2) + \frac{k_\st^{ij\alpha\beta} {S_{TT}}_{\alpha\beta}}{M^4} \,h_{1TT}^{\perp\perp}(x,\bm{k}_\st^2) \right] .
\end{align}

With respect to the original nomenclature in ref.~\cite{Meissner:2007rx}, we have defined $g_1 \equiv g_{1L}$, as well as
\begin{align}
    h_1(x,\bm{k}_\st^2) &\equiv h_{1T}(x,\bm{k}_\st^2) + \frac{\bm{k}_\st^2}{2M^2} \,h_{1T}^\perp(x,\bm{k}_\st^2) ,
\end{align}
as is explained in detail in ref.~\cite{Boer:2016xqr}.

\section{Correlators in coordinate space}
The correlators that we have introduced in the previous section can be translated to $b_\st$-space, where $b_\st$ is Fourier conjugate to $k_\st$. Let us denote a generic TMD correlator by $\chi(x,\bm{k}_\st)$ and a generic TMD function by $f(x,\bm{k}_\st^2)$. Their counterparts in $b_\st$-space are related by a Fourier transformation \cite{Boer:2011xd,Boer:2016xqr}:
\begin{align}
    \tilde{\chi}(x,\bm{b}_\st) &\equiv \int d^2\bm{k}_\st \;e^{i \bm{k}_\st \cd \bm{b}_\st} \,\chi(x,\bm{k}_\st) , \label{e:FT} \\
    \tilde{f}(x,\bm{b}_\st^2) &\equiv \int d^2\bm{k}_\st \;e^{i \bm{k}_\st \cd \bm{b}_\st} \,f(x,\bm{k}_\st^2) . \label{e:FT2}
\end{align}

In momentum space, the parametrization of the correlator in terms of TMDs $f_j$ with rank $n = n(j) \geq 0$, takes the form\footnote{For simplicity, possible Lorentz indices on $\chi$ or $C_j$ are omitted.}
\begin{equation}
    \chi(x,\bm{k}_\st) = \sum_j C_j \,\frac{k_\st^{i_1 \ldots i_n}}{M^n} \,f_j(x,\bm{k}_\st^2) ,
    \label{e:mom space}
\end{equation}
where $C_j$ is a coefficient independent of $k$ that contains information on both the hadron and parton polarization. Using eqs.~\eqref{e:STT}, \eqref{e:FT}, and \eqref{e:mom space}, the correlator in $b_\st$-space is given by
\begin{align}
    \tilde{\chi}(x,\bm{b}_\st) &= \sum_{j} C_j \int d^2\bm{k}_\st \;e^{i \bm{k}_\st \cd \bm{b}_\st} \,\frac{k_\st^{i_1 \ldots i_n}}{M^n} \,f_j(x,\bm{k}_\st^2) \nn \\
    &= \sum_{j} C_j \,\frac{b_\st^{i_1 \ldots i_n}}{M^n} \int_0^\infty d|\bm{k}_\st| \,|\bm{k}_\st| \left( \frac{|\bm{k}_\st|}{|\bm{b}_\st|} \right)^n (2\pi i^n) \,J_n(|\bm{k}_\st||\bm{b}_\st|) \,f_j(x,\bm{k}_\st^2) \nn \\
    &= \sum_{j} \frac{i^n}{n!} \,C_j \,M^n \,b_\st^{i_1 \ldots i_n} \,\tilde{f}_j^{(n)}(x,\bm{b}_\st^2) ,
    \label{e:corrinbtspace}
\end{align}
where on the second line the Bessel function of the first kind $J_k(z)$ arose from the integral identity
\begin{equation}
    \int_0^{2\pi} d\alpha \;e^{ik\alpha} e^{iz\cos(\alpha-\beta)} = 2\pi i^k J_k(z) \,e^{ik\beta} .
\end{equation} 
Following the conventions in refs.~\cite{Boer:2011xd,Boer:2016xqr}, we define the function $\tilde{f}^{(n)}$ as
\begin{align}
    \tilde{f}^{(n)}(x,\bm{b}_\st^2) &\equiv n! \left( - \frac{2}{M^2} \frac{\p}{\p \bm{b}_\st^2} \right)^n \tilde{f}(x,\bm{b}_\st^2) \nn \\
    &= \frac{2\pi n!}{M^{2n}} \int_0^\infty d|\bm{k}_\st| \,|\bm{k}_\st| \left( \frac{|\bm{k}_\st|}{|\bm{b}_\st|} \right)^n J_n(|\bm{k}_\st||\bm{b}_\st|) \,f(x,\bm{k}_\st^2) ,
    \label{e:ftilde}
\end{align}
where we used eq.~\eqref{e:FT2} as well as the recurrence relation
\begin{equation}
    \left( \frac{1}{z} \frac{d}{dz} \right)^m \left[ \frac{J_k(z)}{z^k} \right] = (-1)^m \,\frac{J_{k+m}(z)}{z^{k+m}} ,
\end{equation}
with $z = |\bm{k}_\st||\bm{b}_\st|$ (with $|\bm{k}_\st|$ fixed), $m = n$, and $k = 0$. 

From eqs.~\eqref{e:corrinbtspace} and~\eqref{e:ftilde} it follows that in the parametrization of the correlator in $b_\st$-space the $n$th derivative $\tilde{f}^{(n)}$ with respect to $\bm{b}_\st^2$ appears rather than the function $\tilde{f} = \tilde{f}^{(0)}$ itself. Furthermore, we infer from eq.~\eqref{e:ftilde} that for definite-rank TMDs there is a one-to-one correspondence between the functions in momentum and coordinate space. The motivation for this particular definition of $\tilde{f}^{(n)}$ in eq.~\eqref{e:ftilde} becomes obvious once we set $\bm{b}_\st = 0$, which is equivalent to integration over transverse momentum in $k_\st$-space. Using the limit
\begin{equation}
    \lim_{z\to0} \frac{J_k(z)}{z^k} = \frac{1}{2^k k!} ,
\end{equation}
we see that
\begin{equation}
    \lim_{|\bm{b}_\st| \to 0} \tilde{f}^{(n)}(x,\bm{b}_\st^2) = \int d^2\bm{k}_\st \left( \frac{\bm{k}_\st^2}{2M^2} \right)^n f(x,\bm{k}_\st^2) ,
\end{equation}
which is precisely the conventional $n$th moment $f^{(n)}(x)$ of the TMD. Hence, by construction the derivatives in $b_\st$-space are directly related to moments in $k_\st$-space.

\subsection{The quark-quark TMD correlator}
We can use eq.~\eqref{e:corrinbtspace} to translate the quark-quark TMD correlator in eq.~\eqref{e:phiquark} to $b_\st$-space. In ref.~\cite{Boer:2011xd} this was already done for spin-$1/2$ hadrons. For the spin-$1$ case we have
\begin{align}
    \tilde{\Phi}(x,\bm{b}_\st) =& \;\tilde{\Phi}_U(x,\bm{b}_\st) + \tilde{\Phi}_L(x,\bm{b}_\st) + \tilde{\Phi}_T(x,\bm{b}_\st) + \tilde{\Phi}_{LL}(x,\bm{b}_\st) + \tilde{\Phi}_{LT}(x,\bm{b}_\st) + \tilde{\Phi}_{TT}(x,\bm{b}_\st) ,
\end{align}
where
\begin{align}
    \tilde{\Phi}_U(x,\bm{b}_\st) &= \frac{1}{2} \left[ \slashed{\bar{n}} \,\tilde{f}_1(x,\bm{b}_\st^2) + iM \sigma_{\mu\nu} b_\st^\mu \bar{n}^\nu \,\tilde{h}_1^{\perp(1)}(x,\bm{b}_\st^2) \right] , \\[5pt]
    \tilde{\Phi}_L(x,\bm{b}_\st) &= \frac{1}{2} \left[ \gamma^5 \slashed{\bar{n}} \,S_L \,\tilde{g}_1(x,\bm{b}_\st^2) - M \sigma_{\mu\nu} \gamma^5 \bar{n}^\mu b_\st^\nu S_L \,\tilde{h}_{1L}^{\perp(1)}(x,\bm{b}_\st^2) \right] , \\[5pt]
    \tilde{\Phi}_T(x,\bm{b}_\st) &= \frac{1}{2} \left[ iM \slashed{\bar{n}} \,\epsilon_\st^{S_\st b_\st} \,\tilde{f}_{1T}^{\perp(1)}(x,\bm{b}_\st^2) + iM \gamma^5 \slashed{\bar{n}} \,\bm{b}_\st \cd \bm{S}_\st \,\tilde{g}_{1T}^{(1)}(x,\bm{b}_\st^2) \vphantom{\frac{iM^2 \sigma_{\mu\nu} \gamma^5 \bar{n}^\mu b_\st^{\nu\rho} {S_\st}_\rho}{2}} \right. \nn \\
    &\quad\quad\;\, \left. + \,i\sigma_{\mu\nu} \gamma^5 \bar{n}^\mu S_\st^\nu \,\tilde{h}_1(x,\bm{b}_\st^2) + \frac{iM^2 \sigma_{\mu\nu} \gamma^5 \bar{n}^\mu b_\st^{\nu\rho} {S_\st}_\rho}{2} \,\tilde{h}_{1T}^{\perp(2)}(x,\bm{b}_\st^2) \right] , \\[5pt]
    \tilde{\Phi}_{LL}(x,\bm{b}_\st) &= \frac{1}{2} \left[ \slashed{\bar{n}} \,S_{LL} \,\tilde{f}_{1LL}(x,\bm{b}_\st^2) + iM \sigma_{\mu\nu} b_\st^\mu \bar{n}^\nu S_{LL} \,\tilde{h}_{1LL}^{\perp(1)}(x,\bm{b}_\st^2) \right] , \\[5pt]
    \tilde{\Phi}_{LT}(x,\bm{b}_\st) &= \frac{1}{2} \left[ iM \slashed{\bar{n}} \,\bm{b}_\st \cd \bm{S}_{LT} \,\tilde{f}_{1LT}^{(1)}(x,\bm{b}_\st^2) + iM \gamma^5 \slashed{\bar{n}} \,\epsilon_\st^{S_{LT} b_\st} \,\tilde{g}_{1LT}^{(1)}(x,\bm{b}_\st^2) \vphantom{\frac{M^2 \sigma_{\mu\nu} \bar{n}^\nu b_\st^{\mu\rho} {S_{LT}}_\rho}{2}} \right. \nn \\
    &\quad\quad\;\, \left. + \,\sigma_{\mu\nu} \bar{n}^\nu S_{LT}^\mu \,\tilde{h}_{1LT}(x,\bm{b}_\st^2) + \frac{M^2 \sigma_{\mu\nu} \bar{n}^\nu b_\st^{\mu\rho} {S_{LT}}_\rho}{2} \,\tilde{h}_{1LT}^{\perp(2)}(x,\bm{b}_\st^2) \right] , \\[5pt]
    \tilde{\Phi}_{TT}(x,\bm{b}_\st) &= \frac{1}{2} \left[ - \,\frac{M^2 \slashed{\bar{n}} \,b_\st^{\mu\nu} {S_{TT}}_{\mu\nu}}{2} \,\tilde{f}_{1TT}^{(2)}(x,\bm{b}_\st^2) + \frac{M^2 \gamma^5 \slashed{\bar{n}} \,{\epsilon_\st}_{\mu\nu} b_\st^{\mu\rho} {S_{TT}^\nu}_\rho}{2} \,\tilde{g}_{1TT}^{(2)}(x,\bm{b}_\st^2) \right. \nn \\
    &\quad\quad\;\; \left. - \,iM \sigma_{\mu\nu} \bar{n}^\nu b_\st^\rho {S_{TT}^\mu}_\rho \,\tilde{h}_{1TT}^{(1)}(x,\bm{b}_\st^2) - \frac{iM^3 \sigma_{\mu\nu} \bar{n}^\nu b_\st^{\mu\rho\sigma} {S_{TT}}_{\rho\sigma}}{6} \,\tilde{h}_{1TT}^{\perp(3)}(x,\bm{b}_\st^2) \vphantom{\frac{M^2 \gamma^5 \slashed{\bar{n}} \,{\epsilon_\st}_{\mu\nu} b_\st^{\rho\mu} {S_{TT}^\nu}_\rho}{2}} \right] .
\end{align}
The quark TMDs in $b_\st$-space are one-to-one related to their $k_\st$-space counterparts through eq.~\eqref{e:ftilde}.

\subsection{The gluon-gluon TMD correlator}
We can use eq.~\eqref{e:corrinbtspace} to translate the gluon-gluon TMD correlator in eq.~\eqref{e:gammagluon} to $b_\st$-space. It is given by~\cite{Boer:2016xqr}
\begin{align}
    \tilde{\Gamma}^{ij}(x,\bm{b}_\st) =& \;\tilde{\Gamma}_U^{ij}(x,\bm{b}_\st) + \tilde{\Gamma}_L^{ij}(x,\bm{b}_\st) + \tilde{\Gamma}_T^{ij}(x,\bm{b}_\st) + \tilde{\Gamma}_{LL}^{ij}(x,\bm{b}_\st) + \tilde{\Gamma}_{LT}^{ij}(x,\bm{b}_\st) + \tilde{\Gamma}_{TT}^{ij}(x,\bm{b}_\st) ,
\end{align}
where
\begin{align}
    \tilde{\Gamma}_U^{ij}(x,\bm{b}_\st) &= \frac{xP^+}{2} \left[ - \,g_\st^{ij} \,\tilde{f}_1(x,\bm{b}_\st^2) - \frac{M^2 b_\st^{ij}}{2} \,\tilde{h}_1^{\perp(2)}(x,\bm{b}_\st^2) \right] , \\[5pt]
    \tilde{\Gamma}_L^{ij}(x,\bm{b}_\st) &= \frac{xP^+}{2} \left[ i \epsilon_\st^{ij} S_L \,\tilde{g}_1(x,\bm{b}_\st^2) - \frac{M^2 {\epsilon_\st^{\{i}}_\alpha b_\st^{j\}\alpha} S_L}{4} \,\tilde{h}_{1L}^{\perp(2)}(x,\bm{b}_\st^2) \right] , \\[5pt]
    \tilde{\Gamma}_T^{ij}(x,\bm{b}_\st) &= \frac{xP^+}{2} \left[ - \,iM g_\st^{ij} \epsilon_\st^{S_\st b_\st} \,\tilde{f}_{1T}^{\perp(1)}(x,\bm{b}_\st^2) - M \epsilon_\st^{ij} \bm{b}_\st \cd \bm{S}_\st \,\tilde{g}_{1T}^{(1)}(x,\bm{b}_\st^2) \vphantom{\frac{iM^3 {\epsilon_\st^{\{i}}_\alpha b_\st^{j\}\alpha S_\st}}{12}} \right. \nn \\
    &\qquad\qquad\, \left. - \,\frac{iM \big( \epsilon_\st^{b_\st\{i} S_\st^{j\}} + \epsilon_\st^{S_\st\{i} b_\st^{j\}} \big)}{4} \,\tilde{h}_1^{(1)}(x,\bm{b}_\st^2) + \frac{iM^3 {\epsilon_\st^{\{i}}_\alpha b_\st^{j\}\alpha S_\st}}{12} \,\tilde{h}_{1T}^{\perp(3)}(x,\bm{b}_\st^2) \right] , \\[5pt]
    \tilde{\Gamma}_{LL}^{ij}(x,\bm{b}_\st) &= \frac{xP^+}{2} \left[ - \,g_\st^{ij} S_{LL} \,\tilde{f}_{1LL}(x,\bm{b}_\st^2) - \frac{M^2 b_\st^{ij} S_{LL}}{2} \,\tilde{h}_{1LL}^{\perp(2)}(x,\bm{b}_\st^2) \right] , \\[5pt]
    \tilde{\Gamma}_{LT}^{ij}(x,\bm{b}_\st) &= \frac{xP^+}{2} \left[ - \,iM g_\st^{ij} \bm{b}_\st \cd \bm{S}_{LT} \,\tilde{f}_{1LT}^{(1)}(x,\bm{b}_\st^2) - M \epsilon_\st^{ij} \epsilon_\st^{S_{LT} b_\st} \,\tilde{g}_{1LT}^{(1)}(x,\bm{b}_\st^2) \vphantom{\frac{iM^3 b_\st^{ij\alpha} {S_{LT}}_\alpha}{6}} \right. \nn \\
    &\qquad\qquad\, \left. + \,iM S_{LT}^{\{i} b_\st^{j\}} \,\tilde{h}_{1LT}^{(1)}(x,\bm{b}_\st^2) - \frac{iM^3 b_\st^{ij\alpha} {S_{LT}}_\alpha}{6} \,\tilde{h}_{1LT}^{\perp(3)}(x,\bm{b}_\st^2) \right] , \\[5pt]
    \tilde{\Gamma}_{TT}^{ij}(x,\bm{b}_\st) &= \frac{xP^+}{2} \left[ \frac{M^2 g_\st^{ij} b_\st^{\alpha\beta} S_{TT\alpha\beta}}{2} \,\tilde{f}_{1TT}^{(2)}(x,\bm{b}_\st^2) - \frac{i M^2 \epsilon_\st^{ij} {\epsilon^{\beta}_\st}_\gamma b_\st^{\gamma\alpha} S_{TT\alpha\beta}}{2} \,\tilde{g}_{1TT}^{(2)}(x,\bm{b}_\st^2) \right. \nn \\
    &\qquad\qquad\, \left. + \,S_{TT}^{ij} \,\tilde{h}_{1TT}(x,\bm{b}_\st^2) - \frac{M^2 {S_{TT}^{\{i}}_\alpha b_\st^{j\}\alpha}}{2} \,\tilde{h}_{1TT}^{\perp(2)}(x,\bm{b}_\st^2) + \frac{M^4 b_\st^{ij\alpha\beta} {S_{TT}}_{\alpha\beta}}{24} \,\tilde{h}_{1TT}^{\perp\perp(4)}(x,\bm{b}_\st^2) \right] .
\end{align}
The gluon TMDs in $b_\st$-space are one-to-one related to their $k_\st$-space counterparts through eq.~\eqref{e:ftilde}.

\section{Overview}
In this section we give a brief overview of all leading-twist TMDs that appear in the study of spin-$0$, spin-$1/2$, and spin-$1$ hadrons. In table~\ref{t:overview_tmds} we have organized the quark and gluon TMDs from eqs.~\eqref{e:phiquark} and~\eqref{e:gammagluon} by hadron and parton polarizations. Of the eighteen different quark TMDs, nine are odd under time reversal ($T$), whereas in the gluon case six out of nineteen functions are $T$-odd.\\

\begin{table*}[!htb]
\begin{center}
{\renewcommand{\arraystretch}{1.4}
\begin{tabular}{|c|c|c|c|}
\hline
\textbf{Quarks} & $\gamma^+$ & $\gamma^+ \gamma^5$ & $i \sigma^{i+} \gamma^5$ \\ \hline \hline
U & $\bm{f_1}$ & & $\color{red}h_1^\perp$ \\ \hline
L & & $\bm{g_1}$ & $h_{1L}^\perp$ \\ \hline
T & $\color{red}f_{1T}^\perp$ & $g_{1T}$ & $\bm{h_1}$, $h_{1T}^\perp$ \\ \hline
LL & $\bm{f_{1LL}}$ & & $\color{red}h_{1LL}^\perp$ \\ \hline
LT & $f_{1LT}$ & $\color{red}g_{1LT}$ & $\color{red}h_{1LT}$, $\color{red}h_{1LT}^\perp$ \\ \hline
TT & $f_{1TT}$ & $\color{red}g_{1TT}$ & $\color{red}h_{1TT}$, $\color{red}h_{1TT}^\perp$ \\
\hline
\end{tabular}} \qquad
{\renewcommand{\arraystretch}{1.4}
\begin{tabular}{|c|c|c|c|}
\hline
\textbf{Gluons} & $-g_\st^{ij}$ & $i\epsilon_\st^{ij}$ & $k_\st^i$, $k_\st^{ij}$, etc. \\ \hline \hline
U & $\bm{f_1}$ & & $h_1^\perp$ \\ \hline
L & & $\bm{g_1}$ & $\color{red}h_{1L}^\perp$ \\ \hline
T & $\color{red}f_{1T}^\perp$ & $g_{1T}$ & $\color{red}h_1$, $\color{red}h_{1T}^\perp$ \\ \hline
LL & $\bm{f_{1LL}}$ & & $h_{1LL}^\perp$ \\ \hline
LT & $f_{1LT}$ & $\color{red}g_{1LT}$ & $h_{1LT}$, $h_{1LT}^\perp$ \\ \hline
TT & $f_{1TT}$ & $\color{red}g_{1TT}$ & $\bm{h_{1TT}}$, $h_{1TT}^\perp$, $h_{1TT}^{\perp\perp}$ \\
\hline
\end{tabular}} 
\caption{An overview of the leading-twist quark and gluon TMDs for unpolarized (U), vector polarized (L or T), and tensor polarized (LL, LT, or TT) hadrons. The functions indicated in boldface also occur as collinear PDFs, and the ones in red are $T$-odd. The Dirac structures $\gamma^+$, $\gamma^+ \gamma^5$, and $i \sigma^{i+} \gamma^5 = \tfrac{1}{2} [\gamma^+,\gamma^i] \gamma^5$ correspond to unpolarized, longitudinally polarized, and transversely polarized quarks respectively, whereas the Lorentz structures $-g_\st^{ij}$, $i\epsilon_\st^{ij}$, and $k_\st^i$, $k_\st^{ij}$, etc. correspond to unpolarized, circularly polarized, and linearly polarized gluons respectively.}
\label{t:overview_tmds}
\end{center}
\end{table*}

An additional benefit of using definite-rank TMDs is that the functions also appearing in the collinear case (i.e.\ those that survive integration over $k_\st$) are simply the rank-$0$ functions. However, for the quark TMDs there is one exception to this, namely the rank-$0$ Bacchetta function $h_{1LT}$. Since this function is $T$-odd, it cannot exist in the collinear case. Even though it survives integration over $k_\st$, it in fact vanishes due to the combined effect of hermiticity and time reversal constraints~\cite{Signori:2016lvd}. From table~\ref{t:overview_tmds} it follows that for both quarks and gluons there are four TMDs that also have a collinear counterpart.

\section{Conclusion}
The parametrizations of TMD correlators in terms of TMDs are usually studied in momentum space. However, certain QCD techniques are applied to coordinate rather than momentum space, such as the implementation of TMD evolution. Hence, it is important to have a one-to-one correspondence between TMDs in both spaces. This is achieved by ensuring that TMDs are of definite rank, which is accomplished by using symmetric traceless tensors in $k_\st$ in the parametrizations. The TMDs in $b_\st$-space are then related to their counterparts in $k_\st$-space through a transformation involving a Bessel function of which the order is equal to the rank of the TMD. In this proceedings contribution we have provided the leading-twist parametrizations, in both momentum and coordinate space, of the quark-quark and gluon-gluon TMD distribution correlators in terms of quark and gluon TMDs. We have considered hadrons that unpolarized, vector polarized, or tensor polarized, which is relevant for experiments involving spin-$0$, spin-$1/2$, or spin-$1$ hadrons.

\begin{acknowledgments}
The presented work has been performed in collaboration with Dani\"el Boer, Sabrina Cotogno, Piet J. Mulders, Andrea Signori, and Ya-Jin Zhou. This research is part of the research program of the ``Stichting voor Fundamenteel Onderzoek der Materie (FOM)'', which is financially supported by the ``Nederlandse Organisatie voor Wetenschappelijk Onderzoek (NWO)'' as well as the EU FP7 ``Ideas'' programme QWORK (contract no. 320389).
\end{acknowledgments}

\bibliography{literature}

\end{document}